\documentclass[11pt]{article}
\usepackage[utf8]{inputenc}
\usepackage{amsmath}
\usepackage{amsfonts}
\usepackage{amssymb}
\usepackage{graphicx}
\usepackage{pgfplots}
\pgfplotsset{compat=1.17}
\usepackage[margin=1in]{geometry}
\usepackage{times}  
\usepackage{enumitem}  
\usepackage{booktabs}  
\usepackage{array}  
\usepackage{url}

\usepackage{multicol}

\newenvironment{onecolabstract}{%
  \begin{center}
  \textbf{\large Abstract}
  \end{center}
  \parskip 0.5em
}{%
  \par
}

\title{Revisiting Quantum Supremacy: Simulating Sycamore-Class Circuits Using Hybrid CPU/GPU HPC Workloads}
\author{
  Bob Wold \\
  \small Quantum Rings Inc., Broomfield, CO, USA
  \and
  Venkateswaran Kasirajan \\
  \small Quantum Rings Inc., Broomfield, CO, USA
}

\date{}  

\begin{document}

\maketitle

\begin{onecolabstract}
We present a framework for effectively simulating the execution of quantum circuits---originally designed to demonstrate quantum supremacy---using accessible high-performance computing (HPC) infrastructure. Building on prior CPU-only approaches, our pipeline combines a single NVIDIA A100 GPU for quantum state construction followed by $N$ parallel CPU jobs that perform distributed measurement sampling.  We validate the fidelity by simulating the 53-qubit, 14-cycle Sycamore circuit and achieving a linear cross-entropy benchmarking (XEB) score of $0.549$, exceeding the published XEB score of $0.002$ from Google's reference data. We then evaluate execution time performance with the more complex 53-qubit, 20-cycle circuit, completing the full 2.5 million-shot workload over $100$ CPU jobs in $01:15:36$, representing a $6.95 \times 10^{7}$ speedup compared to Google’s original classical estimate.  Further, we show that if $1,000$ CPU jobs were employed, the estimated duration would be approximately $00:17:35$, only 12 minutes slower than the time taken by the original QPU-based experiment.  These results illustrate that ``quantum supremacy'' is not fixed and continues to be a moving target. In addition, hybrid classical-quantum strategies may provide broader near-term quantum utility than once thought.
\end{onecolabstract}

\begin{multicols}{2}

\section{Introduction}
Quantum computing promises a new class of algorithms that can outperform classical computers on specific problems. One of the most publicized milestones in this space came in 2019, when Google claimed to achieve ``quantum supremacy'' by executing a 53-qubit random circuit sampling (RCS) task on its Sycamore processor in 200 seconds--- a task it estimated would take 10,000 years on a classical supercomputer~\cite{google_supremacy_2019} on a million cores. This landmark claim drew global attention but also immediate skepticism, including a counter-analysis that argued that the problem could be simulated classically in under three days using tensor network methods~\cite{ibm_rebuttal_2019}.

Since then, improvements in classical simulation techniques have continued to minimize the time gap between quantum hardware and classical emulation. In 2024, the same circuits were shown to be simulated on a single CPU-only node with modest memory requirements~\cite{quantum_rings_2024}, achieving high fidelity while only requiring 2.5 days of compute time to sample the circuit 2.5 million times.

This work builds on that foundation by introducing a hybrid simulation pipeline that leverages both GPU acceleration and CPU parallelism to reduce execution time while maintaining fidelity. Using a single NVIDIA A100 GPU for quantum state construction and 100 CPU jobs for distributed sampling, we simulate Google's Sycamore benchmark. We also validate fidelity against Google's published reference data for the 14-cycle version of the circuit.

Our results suggest that the boundary of classical intractability is not fixed. With algorithmic innovation and increasingly accessible high-performance computing (HPC) infrastructure, problems once thought to be exclusive to quantum devices are becoming tractable in practice with classical simulation of quantum systems. This challenges rigid interpretations of quantum supremacy and opens the door to exploring other problems assumed to be intractable using a similar hybrid classical-quantum approach.

\section{Background and Related Work}
Google’s quantum supremacy experiment~\cite{google_supremacy_2019} was designed around random circuit sampling (RCS), a task that produces outputs with probability distributions difficult to simulate classically due to quantum interference and entanglement. Google estimated that the largest 53-qubit, 20-cycle circuit could be simulated with 0.1\% fidelity using a hybrid Schrödinger--Feynman algorithm on Google's cloud servers using 50 trillion core-hours, consuming one petawatt hour of energy, whereas it took only 200 seconds to sample the circuit on the quantum processor three million times.

IBM responded with an alternative classical approach using tensor network contractions~\cite{ibm_rebuttal_2019}, estimating that the task could be simulated on the Summit supercomputer in 2.5 days. While they did not carry out these computations, they provided a detailed description of the proposed simulation strategy as well as the time estimation methodology. However, this solution was circuit-specific, highly optimized, lacked generality, and still required substantial compute resources.

Quantum Rings and Arizona State University later demonstrated a full-circuit simulation of task using one CPU-only node with limited memory~\cite{quantum_rings_2024}. These results exceeded Google's reported fidelity using commodity hardware while achieving a similar runtime to IBM's estimate of 2.5 days, using the Quantum Rings simulation engine.

\section{Methods}
\subsection{Circuit Definitions}
To assess fidelity, we simulate the same 53-qubit, 14-cycle random circuit used in Google's original quantum supremacy study, with gate pattern \texttt{EFGH}. This is a configuration for which Google published reference amplitudes, enabling a direct cross-entropy benchmarking (XEB) comparison. Google reported an XEB score of $0.002$ for this circuit using their quantum processor.

To evaluate performance and run-time characteristics, we simulate the most computationally complex configuration published in the same dataset: a 53-qubit, 20-cycle circuit with gate pattern \texttt{ABCDCDAB}. This circuit is the most complex of the published circuits, and is the circuit used to demonstrate the supremacy claim.

\subsection{Execution Pipeline}
Our execution pipeline is divided into four stages optimized for GPU/CPU hybrid classical hardware. All circuits are initially constructed from Google's QASM-format files.
\begin{enumerate}[leftmargin=*]
    \item \textbf{Quantum State Construction:} A single NVIDIA A100 GPU is used to construct the complete quantum state from the circuit definition. This process takes approximately six minutes and produces a quantum state saved in Quantum Rings' internal format.
    \item \textbf{Persistence:} The generated quantum state, sized at approximately 304.95 MB, is written to a shared file system. The file is optimized for fast read access by the sampling jobs.
    \item \textbf{Distributed Sampling:} Upon completion of quantum state generation, $N$ CPU-only jobs---each provisioned with 16 GB of RAM and 8 CPU cores---are triggered automatically. Each job rebuilds the quantum state from the persisted state and performs $2.5\times10^6/N$ measurement shots. Sampling jobs are implemented as Python scripts using the Quantum Rings SDK. Each job stores its output in a job-specific file using its SLURM Job ID and logs measurement amplitudes and performance metrics.
    \item \textbf{Post-Processing:} A consolidation script aggregates the results across all jobs. This step computes output distributions, aggregates telemetry, and calculates the linear cross-entropy benchmarking (XEB) score.
\end{enumerate}
All stages are orchestrated using SLURM. Jobs are independent and fault-tolerant, supporting opportunistic scheduling and robust execution at scale.

\section{Experimental Setup}
All experiments were conducted on the Sol supercompute hosted by Arizona State University~\cite{HPC:ASU23}.

\subsection{Hardware Configuration}
Sol is a large-scale, production-grade academic HPC resource designed to support a broad range of computational research domains.

\textbf{NOTE:} While this system contains substantial resources, this experiment was performed with just a small subset of these resources.

The Sol supercomputer is comprised of:
\begin{itemize}[leftmargin=*]
    \item \textbf{Compute Cores:} 18,000 CPU cores across 178 compute nodes.
    \item \textbf{Memory:} Standard nodes feature 512 GB RAM; high-memory nodes support up to 2 TB RAM.
    \item \textbf{GPU Resources:} 290 NVIDIA A100 GPUs.
\end{itemize}
Our experiments used a single A100 GPU node for quantum state construction and $N$ CPU-only SLURM jobs for distributed measurement sampling with values of $N$ ranging from 50 to 1,000.

\subsection{Software Environment}
The system runs a standard Linux environment with the following configuration:
\begin{itemize}[leftmargin=*]
    \item \textbf{Operating System:} Rocky Linux 8.10 (HPC-tuned)
    \item \textbf{Scheduler:} SLURM orchestration
    \item \textbf{Quantum Simulation Framework:} Quantum Rings SDK via Python scripts
    \item \textbf{Languages and Tools:} Python 3.11 and Pandas for post-processing.
\end{itemize}

\subsection{Job Scheduling and Execution}
Our hybrid workflow was orchestrated using SLURM job dependencies, Specifically:
\begin{itemize}[leftmargin=*]
    \item A single GPU job was launched to perform quantum state construction.
    \item Upon completion, $N$ parallel CPU-only jobs would be orchestrated through SLURM.
    \item Each CPU job ran independently without inter-node communication.
    \item Jobs were executed under open, opportunistic scheduling windows.
\end{itemize}
Because only a very limited amount of compute time is on a higher end GPU system, and the parallelization is done on CPU only nodes, this demonstrates that the experiment--- which results in a high-fidelity, large-qubit quantum simulations--- can be executed efficiently on commodity-accessible HPC infrastructure with reproducible results.

\section{Results}
\subsection{Fidelity Verification}
To validate the fidelity of our classical simulation, we replicated the 53-qubit, 14-cycle Sycamore circuit using the gate pattern \texttt{EFGH}---the only configuration for which Google published full reference amplitudes. This enables direct comparison via linear cross-entropy benchmarking (XEB).

Using the Quantum Rings SDK, we performed 2.5 million measurements and computed an XEB score of $0.549$. This exceeds Google's reported XEB of $0.002$ for the same circuit executed on quantum hardware, confirming that our simulation accurately reproduces the expected quantum output distribution.

However, this result is modestly lower than the XEB score of $0.678$ reported in prior Quantum Rings work~\cite{quantum_rings_2024}, and while this is still a strong value, additional research is warranted to isolate the contributing factors.

\subsection{Performance Benchmarking}
To evaluate end-to-end performance, we executed the 53-qubit, 20-cycle Sycamore circuit with gate pattern \texttt{ABCDCDAB} using our hybrid classical simulation pipeline. The experiment involved quantum state construction on one NVIDIA A100 GPU, followed by distributed measurement sampling across 100 CPU jobs.

Because this study was conducted using opportunistic access to the Sol supercomputer, CPU jobs experienced queue delays based on shared cluster availability. As such, the actual run time includes queue times.

The complete experiment was completed in 4,536 seconds (01:15:36) from start to finish.

Relative to Google's classical estimate of 10,000 years (approx. $3.15 \times 10^{11}$ s), this constitutes an effective speedup of over $6.95 \times 10^{7}$. Compared to the previous CPU-only method by Quantum Rings, which took approximately 2.5 days, our pipeline delivers a $47.6\times$ performance improvement while maintaining high fidelity.

The results demonstrate that with innovative simulation technology, modest compute resources, and careful workload decomposition, even circuits previously deemed intractable have now been simulated in just over an hour.

\paragraph{Detailed Timing Statistics}
To better understand the observed delays and job-level variability, we captured queue wait times and sampling durations for all 100 CPU jobs. These are summarized in the following:

\begin{center}
\begin{tabular}{|l|r|r|r|}
\hline
\textbf{Metric} & \textbf{Min (sec)} & \textbf{Max (sec)} & \textbf{Avg (sec)} \\
\hline
State Calc & 748 & 748 & 748 \\
Queue & 12 & 618 & 326 \\
Sampling & 2,591 & 3,469 & 2,850 \\
\hline
\end{tabular}
\end{center}
\noindent \textbf{Table 1:} Job-level queue wait times and sampling durations for 100 CPU jobs. Sampling time is consistent; queue wait time reflects opportunistic scheduling.

\subsection{Scalability Discussion}
The simulation workload consists of two distinct phases with different scaling characteristics. Quantum state construction, performed on a single A100 GPU, is currently not allowed to be parallelized with the Quantum Rings SDK. This step requires approximately six minutes, regardless of job count. Any future reduction in total runtime will require architectural updates or GPU-level parallelism during state construction.

In contrast, the measurement sampling phase scales linearly with the number of CPU jobs. Each job independently reconstructs the persisted quantum state and performs a fixed number of measurement shots---e.g., 25,000 in the 100-job configuration. 

To quantify this, we ran smaller single-job experiments at different shot counts. Table~\ref{tab:scaling_table} reports the runtime observed for one job in each configuration:

\begin{center}
\begin{tabular}{|c|c|c|}
\hline
\textbf{Jobs} & \textbf{Shots Per Job} & \textbf{Sampling Time (sec)} \\
\hline
\textbf{100} & \textbf{25,000} & \textbf{3,032.0165} \\
250 & 10,000 & 1,202.5353 \\
500 & 5,000 & 601.8025 \\
1000 & 2,500 & 306.3081 \\
\hline
\end{tabular}
\end{center}
\label{tab:scaling_table}
\noindent \textbf{Table 2:} Observed runtime of a single job for various shot counts. The bold row indicates the actual experiment configuration.

As shown, sampling time scales approximately linearly with shots per job. Doubling the number of jobs roughly halves the per-job runtime. These results were used to extrapolate the expected total runtime under ideal parallelization.

\begin{center}
\begin{tikzpicture}
\begin{axis}[
    width=0.45\textwidth,
    height=5cm,
    xlabel={Number of CPU Jobs},
    ylabel={Sampling Time (seconds)},
    grid=major,
    legend pos=north east,
    xtick={ 200, 400, 600, 800, 1000},
    xmax=1050,
    xmin=0,
    ytick={0, 500, 1000, 1500, 2000, 2500, 3000},
    ymax=3500,
    ymin=0
]
\addplot[color=blue, mark=square*] coordinates {
    (100, 3032.0165)
    (250, 1202.5353)
    (500, 601.8025)
    (1000, 306.3081)
};
\addlegendentry{Observed Sampling Time}
\end{axis}
\end{tikzpicture}
\end{center}
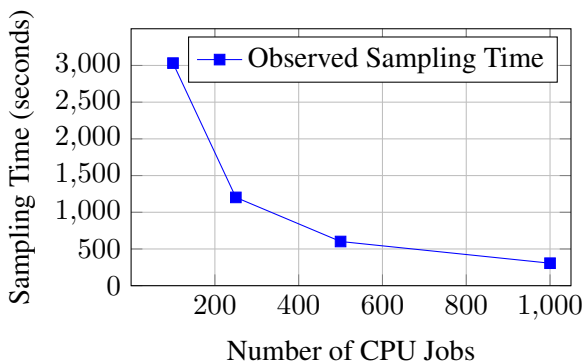
\noindent \textbf{Figure 1:} Observed sampling time for a single job at various levels of parallelism.

Although the complete experiment was executed using 100 parallel CPU jobs, the consistent linear scaling observed in smaller single-job runs provides strong evidence that the approach generalizes to larger configurations. Extrapolating from these measurements, we estimate that total execution time could be reduced to approximately 1,055 seconds (748 seconds for sampling and 306 seconds for quantum state preparation) when scaled to 1,000 jobs.

Of course, real-world performance in HPC environments may vary due to factors such as I/O contention, queue delays, or shared file system load. However, these results demonstrate that the sampling phase can be highly parallelized, offering a practical path to faster execution using standard HPC infrastructure.

\section{Discussion}
This work demonstrates that large-scale quantum circuits---specifically those designed to showcase quantum supremacy---can now be simulated on classical HPC infrastructure with high fidelity and tractable runtime. Although the performance gap between quantum hardware and classical simulation remains real, it is no longer insurmountable at this scale of problem.

By leveraging an execution model that combines GPU acceleration for quantum state construction with CPU-parallel sampling, we have successfully simulated a 53-qubit, 20-cycle Sycamore-class circuit in just over one hour, with the potential to complete in well under an hour with sufficient compute. This challenges the 2019 assertion that such workloads require 10,000 years of classical compute time and suggests that recent improvements in classical simulation tooling and orchestration have meaningfully shifted the quantum-classical boundary.

Importantly, we achieve this result using reasonably available HPC hardware with no privileged access to exotic architectures or proprietary data. The use of public datasets, transparent methodology, and widely available compute resources increases the reproducibility and accessibility of this work.

From a broader perspective, our results suggest that quantum supremacy claims must be understood in context: not as fixed milestones, but as moving benchmarks shaped by improvements on both the quantum and classical sides. This motivates a more nuanced conversation around quantum utility---where hybrid classical-quantum strategies may deliver practical value well before full fault-tolerant quantum systems are deployed.

\section{Conclusion}
Our results demonstrate that quantum circuits previously engineered to be classically intractable are now executable on conventional HPC infrastructure, with practical run times and high fidelity. This not only narrows the performance gap between classical and quantum devices, but also repositions the role of classical simulation in near-term quantum research.

Rather than viewing quantum supremacy as a static boundary, we argue for framing it as a moving frontier---one shaped as much by classical innovation as by quantum hardware advances. In this context, CPU/GPU simulation pipelines can serve both as benchmarks and as practical tools for validating quantum algorithms at scale.

Looking forward, further improvements in state construction efficiency, I/O handling, and job-level scheduling could push this boundary even further. Our work shows that classical simulation remains a living, competitive frontier in quantum computing---and one that continues to advance rapidly.

\section*{Acknowledgments}
We gratefully acknowledge Dr. Gil Speyer and Dr. Torrey Battelle at Arizona State University for providing access to the Sol supercomputer and for their support throughout the execution of this work. We also thank ASU Research Computing for maintaining and enabling academic access to high-performance infrastructure \cite{HPC:ASU23}.

\section*{Reproducibility}
This paper is accompanied by a complete set of source code to reproduce the results. The open source repository is available at \url{https://github.com/Quantum-Rings/quantum-rings-rcs-gpu-cpu}, which includes:
\begin{itemize}[leftmargin=*]
  \item All SLURM job scripts used for orchestration.
  \item All Python scripts used for each phase of simulation.
  \item The QASM files for the 53-qubit, 14-cycle and 10-cycle Sycamore circuits.
  \item The measurement aggregation scripts and post-processing tools used to compute XEB scores.
  \item Instructions for deploying the pipeline on standard Linux HPC systems with SLURM.
\end{itemize}

\bibliographystyle{plain}  
\bibliography{references}

\end{multicols}

\end{document}